# She Innovates- Female owner and firm innovation in India

Shreya Biswas

Assistant Professor, Department of Economics and Finance

Birla Institute of Technology and Science, Pilani, Hyderabad Campus

India

shreya@hyderabad.bits-pilani.ac.in

# She innovates- Female owner and firm innovation in India

*Abstract*

*Using data from World Bank Enterprises Survey-2014, we find that having a female owner in India increases firm's innovation probability using both input and output indicators of innovation. We account for possible endogeneity of female owner variable using a two stage instrumental variable probit model. We find that the positive effect of female owner variable is observed in the sub-samples of firms with more access to internal funding, young firms and firms located in regions with no or less crime This study highlights the need to promote female entrepreneurship as a potential channel for promoting firm innovation in India.*

1. Introduction

India is a lower-middle-income economy as per the World Bank classification and innovation remains a critical tool for growth. The government has long recognized the importance of entrepreneurship and innovation for promoting growth in the country. The Government of India (GoI) has launched the India Innovation Fund in 2007 in a public-private partnership model to mentor innovators from diverse sectors in the country. Recently, the government has also introduced policy initiatives to promote women's entrepreneurship by enabling female entrepreneurs to access cheap capital from formal financial institutions. This paper sheds light on whether promoting female entrepreneurship can also generate benefits for the firm.

The paper considers data from the World Bank Enterprise Survey of 2014 and finds that firms having female owners in India are more likely to innovate both using output and input measures of innovation. Our estimation methodology accounts for endogeneity and the findings are based on two-stage instrumental variable probit models. The sub-sample analysis suggests that innovation is higher among firms with female owners facing lower internal funding constraints. This finding is in line with the literature which suggests that female owners face financing constraints (Cavalluzzo et al., 2002; Farlie and Robb, 2009). When the access to internal funding is low and the firm has to rely on external finance for funding innovation, the positive effect of female owners vanishes. Further, younger firms with female owners are more likely to innovate compared to younger male-owned firms. Young firms with female owners innovate more to possibly reduce the credit constraint in the future and signal its growth and revenue potential to the lenders. Finally, we find that firms with female owners located in regions where either there is no crime or very low crime are more likely to innovate. This highlights the importance of a conducive local business environment on female owner and firm innovation relation.

The contribution of this study is threefold. First, our study extends the debate on how to promote innovation especially in developing economies. Aubert (2005) finds that promoting innovation remains a challenge in developing countries given poor law enforcement, governance and lower human capital owing to the modest educational attainment of people. We find evidence that promoting female entrepreneurship could be one possible channel that can foster innovation among firms in developing countries like India. Second, this study contributes to the literature on female entrepreneurship. There is a growing literature on constraints faced by female entrepreneurs (Muravyev et al., 2009; Farlie and Robb, 2009; Cavalluzzo et al., 2002). The literature on positive effects of having females in decision-making roles has focused on female representation on boards and the performance of large listed firms (Adams and Ferreira, 2009; Khan and Vieito, 2013; Bennouri et al., 2018). The studies on the effect of female ownership on large, medium and small firms are still scant in the context of developing countries. Our paper provides evidence in favor of the positive effect of having a female owner on innovation of firms in India. The results are similar to the findings of Dohse et al. (2019) in the context of European countries.

The rest of the paper is organized as follows. Section 2 provides the background literature on innovation, female owners and firm outcomes, and the theoretical link between gender of the owner and firm-level innovation. Section 3 describes the data and the variables in the study. Section 4 elaborates the methodology employed, while Section 5 presents the results. Finally, Section 6 discusses the implications of the results and concludes.

2. Background literature and research question

2.1 Innovation

Early works by Schumpeter and Arrow referred to either product or process innovation. Product (service) innovation is related to the introduction of a product that is radically different from the existing products or a variant of the existing product in the market. On the other hand, process innovation refers to improvement in the production technology or any other technological development that reduces the marginal cost for the producer. In addition to product and process innovation, organizational and marketing innovation are also two important aspects that have been considered in the literature. Managerial slack can deter firm-level innovation (Rosner, 1968). Innovation diffusion depends upon the organization's ability and willingness to introduce new products. Wang and Ahmed (2004) define organizational innovation as "organization's public ability to introduce new products to market or opening new markets through innovative processes and behaviour". The third edition of the OECD Oslo Manual (2005) defines the four forms of innovation. In addition to these forms of innovation, several studies have considered the input factor i.e. research and development expenditure of the firms as a measure of firm-level innovation.

There is evidence that firm-level innovation is related to improved productivity in developing countries like Pakistan (Wadho and Chaudhry, 2018), among Chinese manufacturing firms (Feng et al., 2019) and firms in Argentina (Chudnovsky et al., 2006). Technological innovation is positively related to the performance of Kenyan firms (Chege et al., 2020). Lee et al. (2019) find that for high technology firms, product innovation is related to superior firm performance and the effect is stronger for firms investing in marketing innovation. On the other hand, for lower technology firms, process innovation improves the performance of firms with organizational innovation. Coad and Rao (2008) find that innovation may not be important for the growth of an average firm; however, it is crucial for high-growth firms. Van Stel et al. (2014) find that innovation is positively related to the survival probability of firms during crisis periods.

Given the various positive spillover effects of innovation, a large body of literature has examined the various determinants of innovation. Market structure is one of the most debated factors affecting innovation. According to Schumpeterian theory, competition is not conducive for innovation. This view suggests that monopoly will induce investment in R&D as they have the resources and the benefits of innovation, which is higher future profits will accrue to the monopolist. On the other hand, according to Arrow (1962), monopoly market structure will deter innovation due to the replacement effect (see Baker, 2007 for the debate on market structure and its impact on innovation). Cohen (1995) does not find any consensus in the empirical literature regarding the relationship between competition in the market and firm innovation. Firm size is another widely accepted determinant of firm innovation (Cohen and Klepper, 1996). Hashi and Stojcic (2013) find that large firms are more likely to innovate, but innovation output falls with firm size. Additionally, the export activity of the firms also increases the innovation probability of firms due to the learning effect (Roger, 2004). Bhattacharya and Bloch (2004) find that size, market structure, and trade share matter for innovation. Access to formal finance is another essential determinant of innovation for firms in developing countries (Ullah, 2019). Many of the firms in developing economies rely on informal finance; however, access to formal financial institutions is crucial for the innovation strategies of firms.

### 2.2 Female owners and firm outcomes

The extant literature suggests that differences in firm outcomes can be explained by differences in the gender of the owner. Studies have found that female-owned enterprises are smaller, have lower survival rate (Farlie and Robb, 2009), have low productivity (Belitski and Desai, 2019), and are also less profitable (Hardy and Kagy, 2018). The adverse outcomes of female-owned firms can be attributed to the additional challenges faced by female owners rather than differences in their abilities. Female owners often face more severe financing constraints compared to male-owned firms. Cavalluzzo et al. (2002) find that female small business owners are more likely to be denied a loan than male owners. Female owned small businesses fail to obtain loans as they have low growth potential and often lack collateral (Coleman, 2002). The gender gap in access to credit is also found for medium enterprises in Sub-Saharan Africa (Hansen and Rand, 2014). Female entrepreneurs also pay higher interest on loans (Muravyev et al., 2009). The financing constraint faced by female owners can be due to discrimination in the credit market or a lower social capital of female owners compared to male owners. The social capital of females tends to be smaller and inferior compared to the social connections of males (McPherson and Lovin, 1982). The limited social capital of females can affect their access to finance (Pham and Talavera, 2018).

Studies in corporate finance have also examined the effect of females in top management teams on various firm outcomes in the context of large listed companies. Adams and Ferreira (2009) find that having female directors on board improves attendance. There is evidence that having females in management teams improves the non-financial performance of large firms in the United States (Chadwik and Dawson, 2018).

### 2.3 Female owners and innovation

The theoretical relation between female ownership and innovation is complex and can be explained using multiple economic theories including discrimination theory and the gender congruity theory, risk-aversion principle, resource dependence theory along with benefits of diversity theory.

The discrimination theory suggests that female owners face discrimination in the credit market and have low access to finance. The lower access to capital can have a detrimental effect on investment in innovation projects. The discrimination can be due to gender congruity, especially in a patriarchal society like India. Social norms dictate that women are involved in unpaid household work like child care, elderly care in joint family and preparing food among others. In patriarchal societies, female entrepreneurs (leaders) can be seen as less favorably than male counterparts (Eagly and Karau, 2002). This can make it more difficult for women to access credit as they are likely to be perceived to be breaking gender stereotypes in society. However, one can argue that female owners are aware of such discrimination and hence, would invest more in innovation projects initially to signal their ability and commitment to the loan approving officer and increase the survival probability and growth potential in the long run. Hence, when there exists gender-based discrimination in credit markets, the effect of gender of the owner on innovation remains ambiguous.

On the other hand, studies on determinants of risk aversion suggest that in general females are more risk-averse compared to males when it comes to investment decisions (Croson and Gneezy, 2009). However, the study did not find differences in risk aversion of males and females in the sample of managers and professionals. Later, Adams and Funk (2012) show that female managers are not representative of the female population and their risk-taking behavior may not be different than male managers. If the risk preferences do not differ between male and female managers, then there should not be any difference in innovation observed for male-owned firms vis-à-vis firms having female owners.

Diversity proponents suggest that having more gender-diverse organizations can improve firm outcomes as men and women face different social, political and economic experiences and such diverse experiences improve group outcomes (Robinson and Denchant, 1997). This idea is closely related to the resource dependence theory (Pfeffer and Slancik, 1978), which suggests that having female owners increases the resources available to the firm, which can improve firm outcomes including innovation. Diverse organizations are found to innovate more in the marketplace (HBR, 2013). Ostergaard et al. (2011) also find that gender-diverse organizations are more likely to innovate compared to organizations dominated by employees of a particular gender. Diaz-Garcia et al. (2013) find that gender-diverse R&D teams come up with superior innovation ideas compared to male-dominated teams.

In the recent past, few studies have examined the relationship between gender of the owner and innovation practices at the firm level. Using a set of emerging countries in Europe, Dohse et al. (2019) and Na and Shin (2019) find that female ownership increases the firm's probability to innovate. Fu et al. (2020) find that female ownership increases the effect of education and experience on product and process innovation among Chinese manufacturing firms. In this study, we examine the role of female owners on both input and output based innovation measures in the context of a patriarchal country like India, where social norms do not favor paid work for women.

India provides an interesting case study for our analysis as it is one of the few countries in the world that has witnessed an overall fall in female labour force participation (Andres et al., 2017) and stagnation in female labour force participation in urban areas (Klasen and Pieters, 2015) in the past few decades. Even though the Indian economy witnessed decent growth since the 1990s, this was not substantiated with an increase in participation of females in the labour force (Lahoti and Swaminathan, 2016). We believe that female owners in India are not representative of the underlying female population and have already broken gender stereotypes by exhibiting entrepreneurial ambitions. These firms with female owners are in a position to reap the benefits of diverse management and decision-making team, which is likely to be conducive for firm-level innovation outcomes. Further, firms having female owners are also likely to face more severe credit constraints. In Indian patriarchal society, female owners will exhibit a tendency to innovate to overcome credit market discrimination and ensure access to finance. The two effects together suggest that having a female owner can be positively related to firm-level innovation in India.

3. **Data and variables**

3.1 **Data**

We consider the World Bank Enterprise Survey India dataset for the year 2014 (WBES). It is a firm-level survey conducted between June 2013 and December 2014. The survey collects data from 9,281 small, medium and large companies using stratified random sampling techniques. The survey provides a snapshot about the business environment in the country. It covers firms operating in the manufacturing, retail and non-retail services industries in 23 states of India. For selecting the manufacturing units, the Annual Survey of Industries (2013) collected by the Central Statistics Office of the Ministry of Statistics and Programme Implementation was used as the sample frame. For services firms, the survey used industry association lists. The data was collected by Nielsen India on behalf of the World Bank and the mode of data collection was face to face interviews. The survey collects information regarding firm characteristics, sales, gender of the owner, innovation and other performance measures.

### 3.2 Variables

We consider both output and input measures of innovation as dependent variables in our study. The output-based innovation measures are given by the enterprise-level process, organizational and marketing innovation indicators. *Product innovation* is a dummy that takes the value one if the firm launched a new product or service during the last three years and zero otherwise. *Process innovation* is a dummy that takes the value one if the firm introduces either (a) a new method of manufacturing or providing services; or (b) improved the logistics and distribution methods; or (c) improved supporting activities like maintenance systems, accounting or computing operations and zero otherwise. *Organizational innovation* is a dummy variable that takes the value one if the firm has introduced any new organizational structure or management practices and zero otherwise. *Marketing innovation* is a dummy that takes the value one if the firm introduced new methods to market its products or services during the last three years and zero otherwise. We consider *R&D* as a proxy for firm-level input measures of innovation. R&D is a dummy that takes the value one if the firm had spent on R&D activities during the last three years or if the firm gives time to its employees to develop or improve new approaches, products or ideas and zero otherwise.

The main interest variable is the female dummy takes the value one if at least one of the owners of the firm is a female and zero otherwise. There are 9,224 firm observations for which information regarding the gender of the owner is available. In our sample, 1,372 firms have at least one female owner.

We control for size, age, export orientation, being a part of a multi-firm establishment and financing constraints faced by the firm. We use the WBES size definition wherein the firms with less than 20 employees are considered small firms, firms with 20-99 employees are considered medium-sized firms and firms with more than 100 employees are considered as large firms. Age of the firm is given by the year of interview minus the year of incorporation. Export orientation of a firm is a dummy if the firm is an exporting firm and zero otherwise. The multi-firm variable is a dummy that takes the value one if the firm is a part of a large multiform establishment and zero otherwise. Financing constraint is a categorical variable that ranges from zero to four, where zero indicates that the firm does not face any difficulty in accessing finance and four indicates that the firm faces a very severe obstacle in accessing finance. We also control for state and 26 industry dummies to control for any state-specific laws and industry-specific regulations or general business environment that can affect the likelihood of firm-level innovation.

4. **Methodology**

Given the binary nature of our dependent variable we estimate probit model to understand the relation between firm-level innovation and female owner dummy. However, it is possible that firms that have female owners are different than the firms without female owners and our estimated $\beta_1$ captures the effect of these differences due to selection bias rather than the effect of female owner on innovation. In order to address the endogeneity related to the female dummy variable we follow a two-stage instrumental variable probit approach. An instrument is a variable that should satisfy the relevance and the exogeneity conditions. The relevance criteria suggest that the instrument is correlated to the female owner variable and the exogeneity condition implies that the instrument should be otherwise uncorrelated to the innovation measures.

We consider the state-industry share of firms that have female owners as an instrument. We consider the theoretical framework suggested by Fisman and Svensson (2007) to construct our instrument. Fisman and Svensson (2007) argue that firms would get engaged in bribery payments either because the firm itself is corrupt or because of average industry factors. Extending a similar argument, we believe the likelihood that the firm will have a female owner can be driven by two factors i.e. specific firm factors and average business environment given by industry and state factors. For example, there are few industries that tend to be favored by females, such as textiles, hospitality etc. Similarly, in India there could be state-specific programs that encourage female entrepreneurship. Hence, a hospitality firm in a state with a policy that encourages female entrepreneurship is more likely to have female owners compared to a mining firm located in a state with no policy targeting female entrepreneurs. Further, a higher share of firms having at least one female owner in a particular industry within a specific industry can also increase the firm *i*'s probability of having a female owner through peer effect. Finally, we believe that the state-industry specific share of firms having female owners is not likely to affect the firm-level strategy to invest in innovation. While calculating the state-industry share of firms having female owners we exclude firm *i*. In the first stage, we regress our female owner dummy on the state-industry share of female firms (instrument), firm controls, industry and state dummies.

$$Female\ owner_{ijs} = \alpha_0 + \alpha_1 state-industry\ share\ of\ female\ owners_i + \partial X_{ijs} +$$
$$\emptyset Indutry_i + \partial State_s + e_{ijs} \qquad (1)$$

Equation (2) is estimated using Ordinary Least Square (OLS) method. When our interest variable is the female dummy, we assume a linear relation in the first stage. In the second stage, we estimate a probit model wherein the fitted values obtained from stage one is included as an explanatory variable along with other covariates given by the equation below:

$$L_{ijs} = b_0 + b_1 \widehat{Female\ owner}_{ijs} + \tau X_{ijs} + \pi Indutry_i + \omega State_s + \epsilon_{ijs} \qquad (2)$$

## 5. Results

### 5.1 Summary statistics

The sample characteristics are presented in Table 1. We consider the strict weight reported by WBES data for computing the sample means. Column1 presents the summary for the full sample. Among the firms for which the product innovation question was answered, 44.9% did introduce a new product or service in the last three years. Around 56.4% firms introduced some process innovation in the past three years preceding the survey. The proportion of firms that have introduced organizational or marketing innovation or engaged in R&D were less than 50%. Only 10.7% firms had female owners and the rest of the firms had male owners.

<<Insert Table 2 here>>

Columns 2-3 summarize characteristics separately for firms that have female owners and those without any female owners. We observe that a higher share of firms with female owners innovate compared to male-owned firms. Female owned firms are also older and are more export oriented. Among the small and medium firms, less than 10% have female owners, whereas for large firms close to 19.5% have female owners. This can be due to the fact that many large firms are part of business groups where the female members from the promoter family could be an owner of the firm. As a result, the share of firms having a female owner seems to be higher for large firms. The univariate statistics suggest that gender of the owner may be related to the innovation practices of firms. However, other firm characteristics also differ across male owned and firms with female owners, we consider a multivariate framework to test our hypothesis.

**5.2 Main results**

Table 2 presents the coefficients of female owner dummy on various firm innovation obtained from probit estimation. We find that the female dummy is unrelated to product innovation (column 1), but is positively related to all other forms of innovation (columns 2-5). Our initial multivariate results indicate there is an association between having a female owner and innovation practice of firms in India.

<<Insert Table 2 here>>

As discussed in the methodology section, probit estimates are likely to be biased if female owner dummy is endogenous. We employ a two-stage IV probit model wherein the instrument is the state-industry share of female owners. Table 3 presents the coefficients from the second stage probit model of innovation on predicted female owner variable obtained from stage one. Female dummy is positive and significant in all the specifications (columns 1-5) for output and input based innovation measures except product innovation. Further, in the first stage regression, the coefficient of state-industry share of female is positive and highly significant in both the specifications suggesting that the instrument is indeed correlated with the endogenous variable. Our results suggest that having a female owner can improve both input and output-based innovation practices for Indian firms. The results are in line with the findings of Dohse et al. (2019) in the context of European countries.

<<Insert Table 3 here>>

### 5.3 Heterogeneous effects

#### 5.3.1 Availability of internal funds

Financing of innovation at the firm level is a major challenge faced by owners and managers. The under-investment in R&D is tied to financing challenges faced by firms. Innovation projects are high-risk projects with low to moderate expected return and a low probability of very high return in the future. The high cost of capital deters firms from investing in innovation projects. Studies suggest that large firms often prefer internal funding for innovation investments (Hall, 2002; Hall and Lerner, 2009).

In line with the financing argument, we separately analyze the effect of having female owner for sub-samples based on the share of working capital and fixed asset expenses funded by internal funds. If the share of working capital and fixed assets that are funded by internal funds is greater than the median level (40%), the firm is considered as less financially constrained; otherwise, it is considered to have access to low or moderate internal financing resources. Figure 1 indicate that the positive effect of female owner on innovation is driven by the sub-sample of firms having sufficient internal funding and female owners do not have any effect on innovation for firms that have low internal funds. This could be due to the fact that firms with female owners with low internal funds have to rely on external funds for funding innovation projects. Several studies highlight that female entrepreneurs face more severe external financing constraints compared to male-owned firms (Asiedu et al., 2013; Sabarwal and Terrell, 2008). Hence, external credit market distortion could be a plausible reason why the positive effect of female ownership vanishes for firms with lower internal funds.

<<Insert Figure 1 here>>

#### 5.3.2 Young versus old

The extant literature suggests that the firm age can be related to innovation. Huergo and Jaumandreu (2004) show that for manufacturing firms, the new entrants are most likely to innovate compared to older firms. Later, Balasubramanian and Lee (2008) find that the age of firms adversely affects the technical innovation of firms given by patents. Younger firms are likely to engage in riskier innovation practices and generate employment (Coad et al., 2016). Other studies have highlighted that age can improve the innovative capacity of the firms. Recently, Petruzzelli et al. (2018) also find that older firms are in a better position to apply mature knowledge and innovate in the market compared to younger firms. However, when it comes to applying new knowledge, the younger firms are in a more beneficial position. Cucculelli, (2018) finds that for Italian firms, firm age is positively related to the probability of product innovation.

We define young firms as those which are in the market for ten years or less and rest are categorized as old firms. Based on this definition, 38% firms in our sample are young. Among young firms, 10% have female owners, while 11% old firms have female owners. We estimate the IV probit models separately for the sample of old and young firms. Figure 2 shows that the positive relation between female owner on innovation is driven by younger firms in the sample. For the sub-sample of older firms, there is no relation between the gender of the owner and likelihood of innovation. It is possible that the femaleinnovate in the initial years to increase their chance of survival and also high innovation in the start-up phase by female owners can improve their credit market outcomes in the later part of the firm's life-cycle.

<<Insert Figure 2 here>>

### 5.3.3 Crime and innovation

Matti and Ross (2016) provide a review of how crime can be related to entrepreneurial activity in various settings. Crime is a cost to business activity (Kuratko et al., 2000) as it increases the perceived loss on account of theft, robbery, etc. Rosenthal and Ross (2010) find that crime in the neighborhood has an impact on location decisions of business in the United States. A higher crime rate in the city tends to deter business activity and entrepreneurs would prefer relocating to safer places to ensure a lower effect on revenues. There is a possibility that high-quality firms self-select to locate themselves in low crime rate locations. These high-quality firms are likely to be innovators. The study by Velasquez (2019) shows that the crime rate has an adverse effect on self-employment and women's labour supply in Mexico. One can argue that if the crime rate is high, the uncertainty in business activity is even higher and this can have a negative effect on investments in high-risk innovation projects. We separately analyze whether the positive effect of female ownership on innovation practices is different for firms based on the crime in the neighborhood.

The WBES specifically asks respondents, "how much of an obstacle is crime, theft and disorder for the business?" If the respondent answers no or minor obstacle for the above question, we consider that the firm is in a low crime region. In case the response is moderate, major and very severe we classify the firm as situated in a high crime region. This suggests the possibility of crime affecting the location decisions of firms. Based on this classification, around 14.5% of firms are in high crime regions and the rest are in low crime regions. Almost 13.25% firms in the low-crime region have female owners and less than 10.5% firms in the high-crime region have female owners. We observe that in the high-crime region, lesser firms are likely to have female owners. A high crime region is likely to reduce physical mobility and hours spent by females outside the household. Figure 3 indicates that female owners increase the likelihood of innovation among firms located in low crime regions; however, there is no effect of female ownership on the probability of innovation for firms in high crime regions. It appears that environmental factors like low crime are important to reap the positive effect of female owner on firm outcomes. Becker and Rubenstein (2011) model of response to crime can be extended to explain the insignificant result of female ownership for high crime sub-sample. The perceived risk of victimization by female owners in high crime regions can negatively affect their involvement in businesses and make them more risk-averse, which in turn may reduce investments in innovation projects.

<<Insert Figure 3 here>>

## 6. Discussions and conclusion

In this study, we find that firms having female owner are more likely to innovate in India after accounting for the possibility of endogeneity. The results are driven by firms with more internal funding resources, younger firms in the sample and firms located in regions with no or low crime. The findings have implications for both future research and policy-making.

The Economic Survey (2020) suggests that entrepreneurship activity is positively correlated with GDP growth in the country, and India ranks third in terms of new firm creation. The GoI initiated the Start-up India initiative in January 2016 to promote entrepreneurship and innovation in the country[1]. Under this initiative, the start-ups are tax exempted for three years. Such policy initiatives are a step in the right direction. Our study suggests that additional initiatives to bridge the gender gap in entrepreneurial activity can serve the twin objectives of empowering women and promoting innovation. The Indian state of Telangana started an incubator in 2018 called WE Hub to promote women entrepreneurship in the state[2]. Additional efforts at the state and national level to formally promote and nurture female entrepreneurs, especially during the first few year post registration are critical for survival and firm-level innovation. Policymakers may consider providing R&D subsidies for start-ups in general and specifically for female owners.

---

[1] https://www.startupindia.gov.in/
[2] http://wehub.telangana.gov.in/

Our study highlights that firms with female owners having access to internal funding are more likely to innovate. It is possible that access to external capital remains a crucial challenge for female owners. There are initiatives like Stree Shakti package[3] and Udyogini scheme[4] that provide access to capital, marketing, training and other support to women entrepreneurs in the country. However, it is imperative to promote such schemes among the target population by conducting roadshows and address social norms to reduce discrimination among women in the credit market. Further, our results also highlight the importance of maintaining good law and order as that has an effect on the business location as well as strategic decisions like innovation investment by female owners. The study provides evidence that promoting female entrepreneurship in developing economies can initiate a virtuous cycle of wealth creation through innovation along with the socially desirable outcome of reducing gender-gap in outcomes.

---

[3] https://sidbi.in/oldsmallb/bank-schemes/stree-shakti-package
[4] http://www.udyogini.org/

**Tables**

Table 1: Summary statistics

The table below presents the summary statistics for innovation outcomes, female ownership variables and other firm characteristics.

|  | Full sample | Firms with female owners | Male-owned firms |
|---|---|---|---|
| *Product innovation* | 0.449 | 0.498** | 0.440 |
| *Process innovation* | 0.564 | 0.610*** | 0.552 |
| *Organizational innovation* | 0.417 | 0.427* | 0.409 |
| *Marketing innovation* | 0.464 | 0.488*** | 0.455 |
| *R&D* | 0.438 | 0.495*** | 0.424 |
| *Female* | 0.107 | 1.000 | ---- |
| *Share of female* | 4.654 | 48.418 | ---- |
| *Size(share):* | | | |
| *Small* | 0.461 | 0.376 | 0.474 |
| *Medium* | 0.404 | 0.387 | 0.409 |
| *Large* | 0.135 | 0.236 | 0.117 |
| *Age(in years)* | 16.157 | 19.485*** | 15.911 |
| *Multifirm(share)* | 0.787 | 0.785 | 0.793*** |
| *Financial Constraint(share):* | | | |
| *No Constraint* | 0.341 | 0.419 | 0.326 |
| *Less Constraint* | 0.333 | 0.247 | 0.345 |
| *Moderate Constraint* | 0.175 | 0.171 | 0.178 |
| *Major Constraint* | 0.106 | 0.108 | 0.108 |
| *Severe Constraint* | 0.044 | 0.056 | 0.044 |
| *Export(share)* | 0.089 | 0.221*** | 0.074 |
| *No. of observations* | 9281 | 1372 | 7852 |

\*\*\* p<0.01, \*\* p<0.05, \* p<0.1

Table 2: Female ownership and innovation- Probit model

The table below presents the coefficients of probit model of innovation outcomes on female owner dummy and other firm characteristics. *** p<0.01, ** p<0.05, * p<0.1

|  | (1) Product innovation | (2) Process innovation | (3) Organizational innovation | (4) Marketing innovation | (5) R&D |
|---|---|---|---|---|---|
| Female owner | 0.080* | 0.159*** | 0.103** | 0.115*** | 0.174*** |
|  | (0.042) | (0.044) | (0.043) | (0.043) | (0.044) |
| Medium | 0.297*** | 0.274*** | 0.127*** | 0.074** | 0.214*** |
|  | (0.034) | (0.034) | (0.035) | (0.035) | (0.036) |
| Large | 0.380*** | 0.383*** | 0.176*** | 0.169*** | 0.323*** |
|  | (0.044) | (0.047) | (0.046) | (0.045) | (0.047) |
| Age | 0.001 | 0.001 | -0.001 | -0.002 | 0.001 |
|  | (0.001) | (0.001) | (0.001) | (0.001) | (0.001) |
| Multi firm | -0.090** | -0.247*** | -0.399*** | -0.318*** | -0.328*** |
|  | (0.038) | (0.040) | (0.039) | (0.039) | (0.040) |
| No/ Less constraint | 0.110*** | 0.296*** | 0.288*** | 0.223*** | 0.323*** |
|  | (0.039) | (0.040) | (0.040) | (0.040) | (0.041) |
| Moderate constraint | -0.045 | 0.244*** | 0.370*** | 0.377*** | 0.432*** |
|  | (0.044) | (0.045) | (0.045) | (0.045) | (0.047) |
| Major constraint | 0.031 | 0.236*** | 0.293*** | 0.355*** | 0.250*** |
|  | (0.056) | (0.056) | (0.056) | (0.056) | (0.059) |
| Severe constraint | -0.166** | 0.008 | 0.089 | 0.074 | 0.012 |
|  | (0.077) | (0.078) | (0.076) | (0.076) | (0.076) |
| Constant | -0.423*** | 0.529*** | 0.591*** | 0.071 | -0.581*** |
|  | (0.090) | (0.097) | (0.097) | (0.093) | (0.093) |
| State FE | Yes | Yes | Yes | Yes | Yes |
| Industry FE | Yes | Yes | Yes | Yes | Yes |
| Observations | 9,037 | 9,032 | 9,029 | 9,029 | 9,004 |

Table 3: IV probit regression output.

The table below presents the coefficients of instrumental variable probit model of innovation outcomes on female owner dummy and other firm characteristics. *** p<0.01, ** p<0.05, * p<0.1

|  | (1) Product innovation | (2) Process innovation | (3) Organizational innovation | (4) Marketing innovation | (5) R&D |
|---|---|---|---|---|---|
| Female owner | 0.024 | 0.155*** | 0.125** | 0.194*** | 0.206*** |
|  | (0.056) | (0.056) | (0.053) | (0.055) | (0.057) |
| Medium | 0.305*** | 0.273*** | 0.122*** | 0.069** | 0.216*** |
|  | (0.034) | (0.035) | (0.035) | (0.035) | (0.037) |
| Large | 0.387*** | 0.378*** | 0.171*** | 0.163*** | 0.319*** |
|  | (0.045) | (0.047) | (0.046) | (0.046) | (0.048) |
| Age | 0.001 | 0.001 | -0.001 | -0.002* | 0.002 |
|  | (0.001) | (0.001) | (0.001) | (0.001) | (0.001) |
| Multi firm | -0.083** | -0.239*** | -0.389*** | -0.306*** | -0.319*** |
|  | (0.038) | (0.041) | (0.039) | (0.039) | (0.041) |
| No/ Less constraint | 0.115*** | 0.293*** | 0.287*** | 0.220*** | 0.316*** |
|  | (0.039) | (0.040) | (0.040) | (0.040) | (0.042) |
| Moderate constraint | -0.044 | 0.234*** | 0.360*** | 0.374*** | 0.425*** |
|  | (0.044) | (0.046) | (0.045) | (0.045) | (0.047) |
| Major constraint | 0.030 | 0.222*** | 0.283*** | 0.349*** | 0.245*** |
|  | (0.056) | (0.056) | (0.057) | (0.057) | (0.059) |
| Severe constraint | -0.173** | 0.007 | 0.090 | 0.076 | 0.011 |
|  | (0.077) | (0.078) | (0.077) | (0.077) | (0.077) |
| State-industry share of females | 0.016*** | 0.016*** | 0.016*** | 0.016*** | 0.016*** |
|  | (0.000) | (0.000) | (0.000) | (0.000) | (0.000) |
| Constant | -0.429*** | 0.533*** | 0.597*** | 0.063 | -0.588*** |
|  | (0.090) | (0.097) | (0.097) | (0.093) | (0.093) |
| State FE | Yes | Yes | Yes | Yes | Yes |
| Industry FE | Yes | Yes | Yes | Yes | Yes |
| Observations | 8,915 | 8,910 | 8,907 | 8,907 | 8,882 |

**Figures**

Figure 1: Female owner and innovation: High versus low internal funding

The figure below presents the coefficients obtained from second stage IV-probit regression of innovation on female owner and other firm characteristics for the sub-sample of firms with low and high internal funding resources.

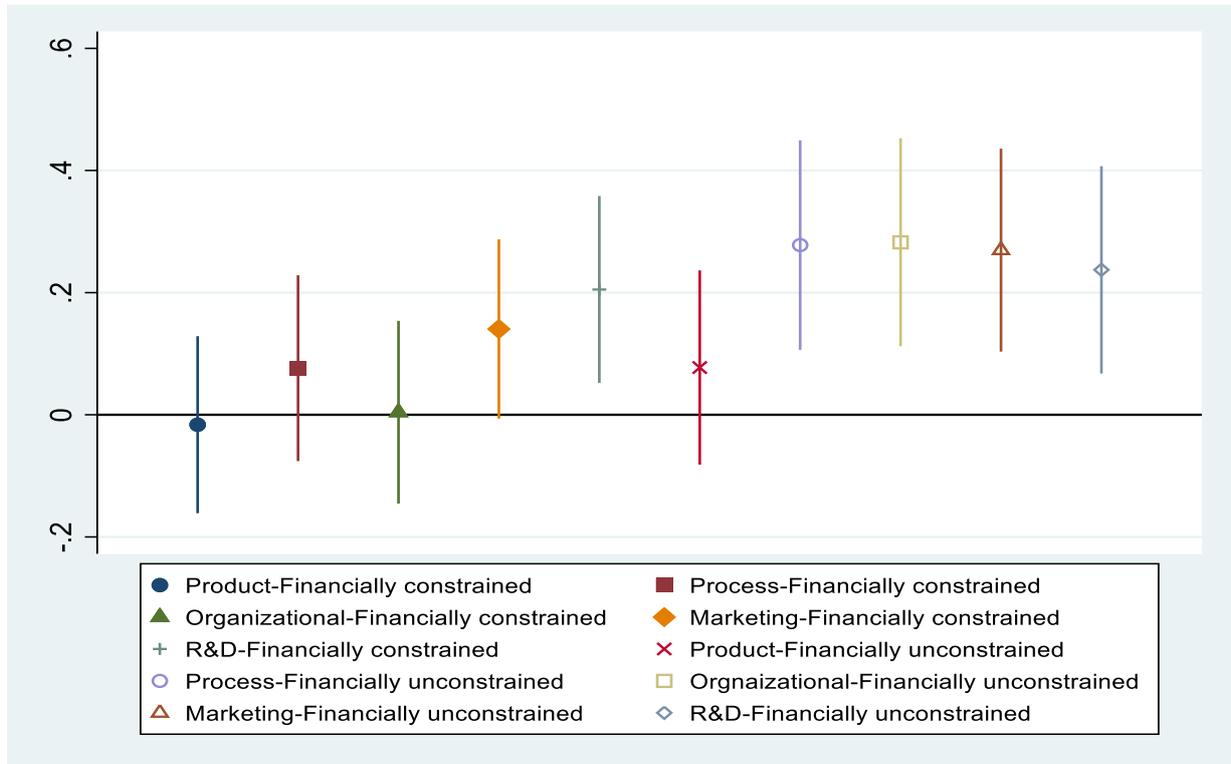

Figure 2: Female owner and innovation: Young versus old firms

The figure below presents the coefficients obtained from second stage IV-probit regression of innovation on female owner and other firm characteristics for the sub-sample of old and young firms in the sample.

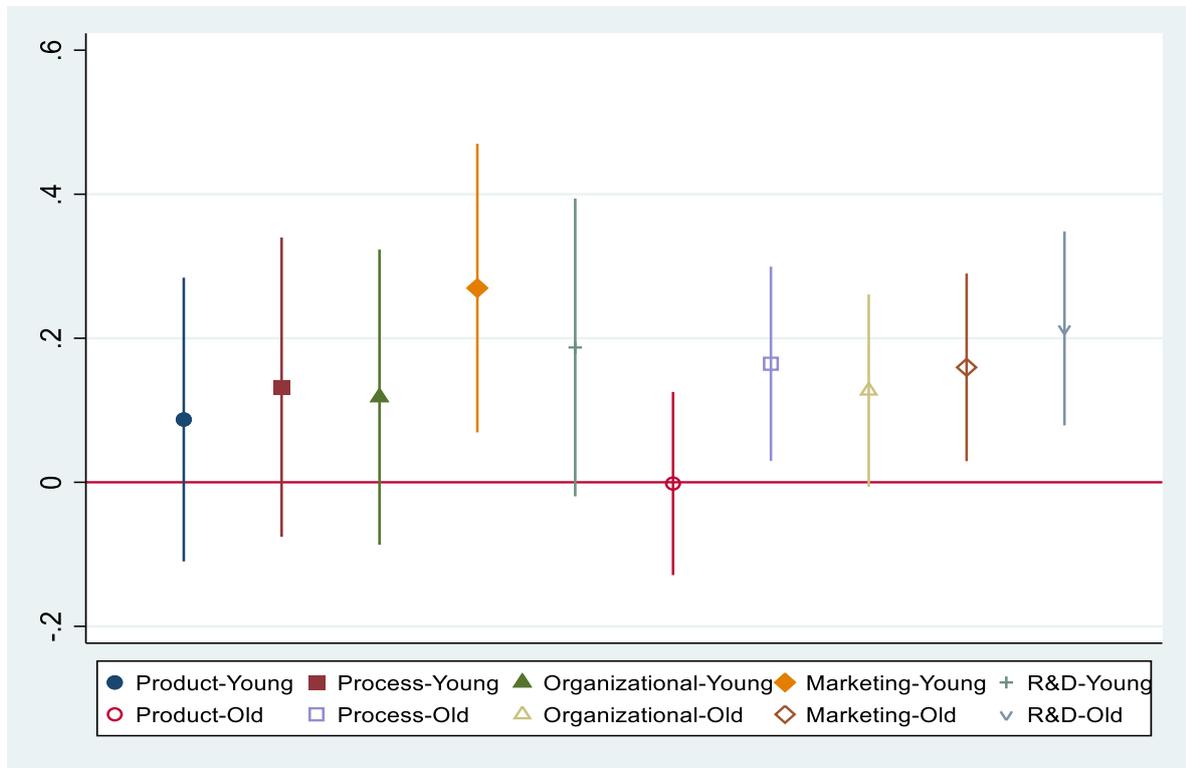

Figure 3: Female dummy and innovation: Low versus high crime region

The figure below presents the coefficient obtained from second stage IV-probit regression of innovation on female owner dummy for the sub-sample of firms in low and high crime region.

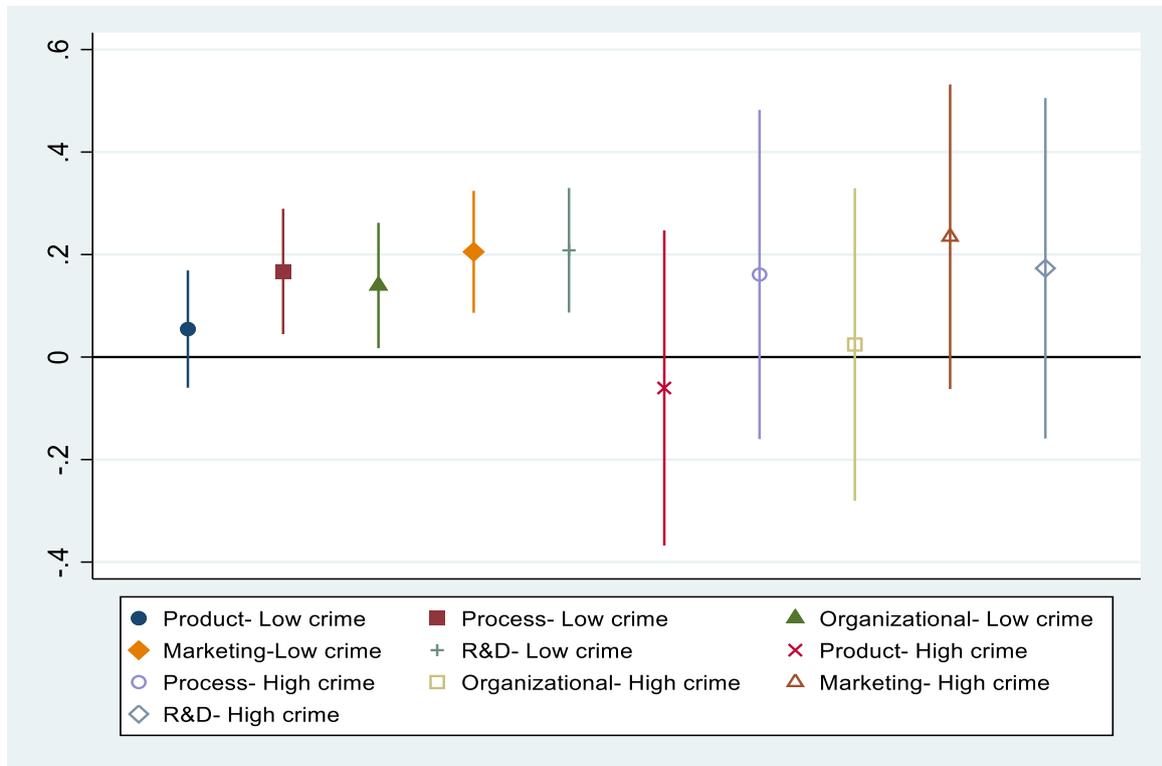